\documentclass[aps,a4paper,twocolumn,10pt,prl]{revtex4-2}
\usepackage[normalem]{ulem}
\usepackage{graphicx}
\usepackage{dcolumn}
\usepackage{bm}
\usepackage{setspace} 
\usepackage{hyperref}

\usepackage[separate-uncertainty=true]{siunitx}[=v2]
\usepackage{scalerel,amssymb}
\usepackage{siunitx}
\usepackage{xcolor}
\DeclareSIUnit[number-unit-product = {}]{\bact}{bact}
\DeclareSIUnit[number-unit-product = {}]{\pixel}{pixel}

\begin{document}

\preprint{APS/123-QED}

\title{Active Saffman--Taylor Viscous Fingering}

\author{Akash Ganesh}
\author{Carine Douarche}%
\author{Harold Auradou}%
 \email{harold.auradou@universite-paris-saclay.fr}
\affiliation{Universit\'{e} Paris-Saclay, CNRS, FAST, 91405, Orsay, France.}

\begin{abstract}
{
Adding swimming bacteria to a liquid causes its effective shear viscosity to decrease, eventually reaching a regime of zero viscosity. 
We examined whether this property leads to viscous finger-like displacement fronts like those observed when a less viscous fluid displaces a more viscous liquid. Our study revealed that this system exhibits more complex dynamic characteristics than the classical Saffman--Taylor instability. We discovered that this instability occurs when the bacterial volume fraction exceeds a critical value, and the imposed shear rate is below critical value, for which the viscosity of the suspension is zero.
}
\end{abstract}
\maketitle
\paragraph{\textbf{Introduction.-} }
An individual bacterium has the ability to convert the chemical energy present in the surrounding fluid into mechanical motion (swimming)~\cite{Lauga2009,Drescher2011}.
From the swimming of a population of bacteria, emerge at large scales, interesting properties of the fluid, such as the appearance of coherent dynamic structures
~\cite{Sokolov2012,Rabani2013,Lushi2014,Gachelin2014}, unidirectional flows
~\cite{Wioland2013,Nishiguchi2018}, enhanced dispersion ~\cite{rusconi_bacterial_2014,vennamneni_shear-induced_2020,ezhilan_transport_2015,Ganesh2023}, active motion of small objects~\cite{Wu2000,Leonardo2010,sokolov_swimming_2010}, dynamical clustering of passive particles
~\cite{Bouvard_Moisy_Auradou_2023} \textit{etc}. Such properties have made bacterial suspensions a popular model system for exploring the physical properties of active fluids~\cite{Ramaswamy_2010,Marchetti2013,Ramaswamy_2017}.
\begin{figure}[hb!]
\centering
  \includegraphics[width=0.5\textwidth]{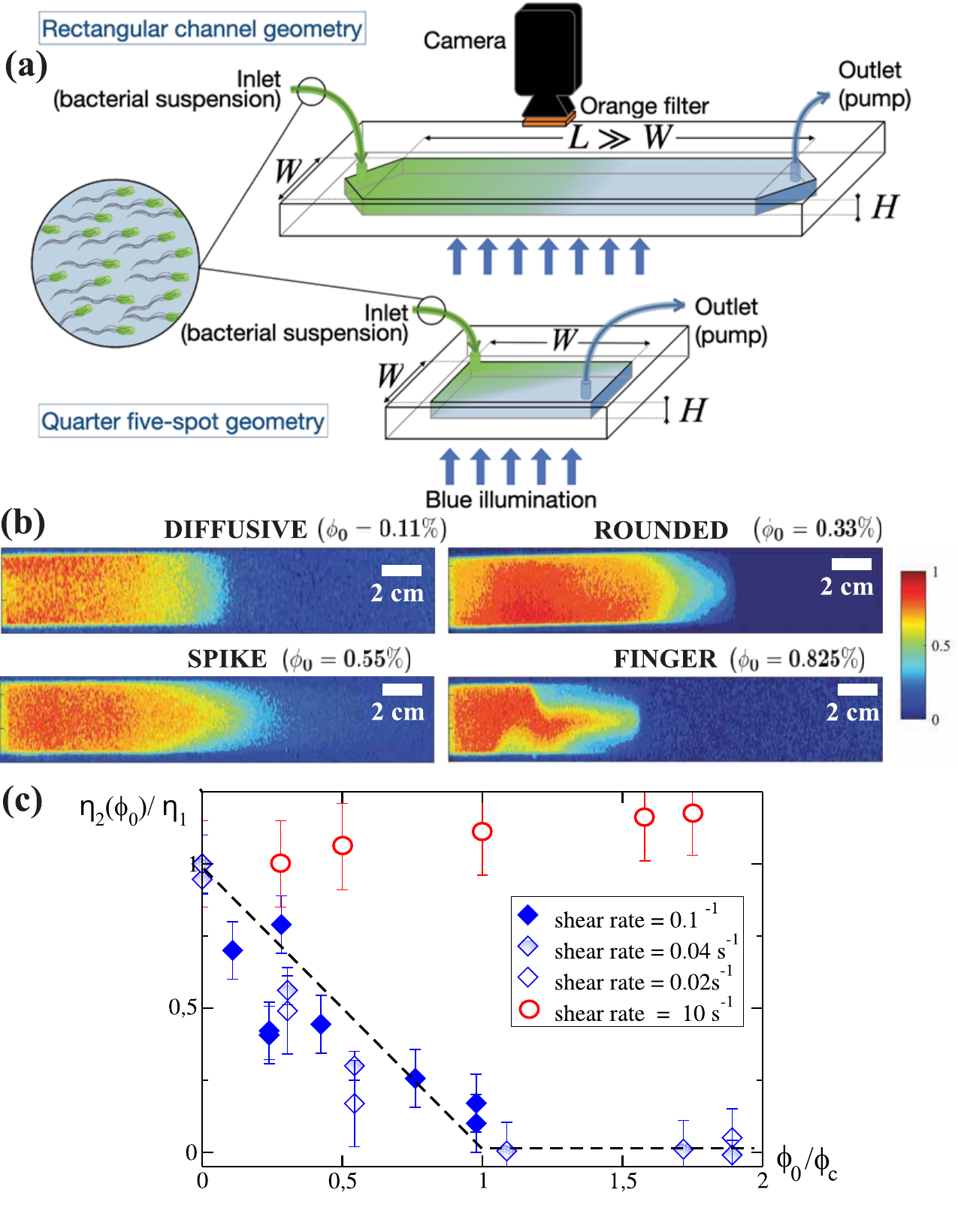}
  \caption{(a) Schematic representations of the HS and QFS geometries.
  Fluorescent bacteria are injected into the aperture initially filled with a buffer and visualized with a camera through a colored filter. HS cells measure $L = 21$~cm in length, $W = 1$ or $2$~cm in width, and $H = 500~\mu$m in aperture (top panel). QFS devices are squares with sides $5$~cm and $H = 500~\mu$m aperture (down panel). (b) Fronts observed for four different bacterial volume fractions $\phi_0$. (c) Viscosities of the bacterial suspensions measured in a Couette rheometer~\cite{lopez_turning_2015,Martinez2020,chui_rheology_2021} at shear rates lower ($\Diamond$) and higher ($\bigcirc$) than the critical shear rate $\dot{\gamma}_c=0.4$~s$^{-1}$ as function of $\phi_0$.   Dashed line: a visual guide for viscosity measurements using the mean-field continuum kinetic theory model~\cite{Martinez2020,saintillan_dilute_2010}.}
  \label{fig1}
\end{figure}
One of the interesting properties observed in a suspension of bacteria is their ability to reduce the effective viscosity of the suspension when it is subject to a moderate shear 
~\cite{hatwalne_rheology_2004,sokolov_reduction_2009,saintillan_dilute_2010,saintillan_extensional_2010,gachelin_non-newtonian_2013,lopez_turning_2015,clement_bacterial_2016,saintillan_rheology_2018, Liu_Zhang_Cheng_2019,Choudhary_Nambiar_Stark_2023,Martinez2020,chui_rheology_2021}. 
This emergence is due to the fact that the spatially organized stress response generated by the bacteria motion aids the applied shear stress, leading to a decrease in the effective viscosity ~\cite{saintillan_dilute_2010,saintillan_rheology_2018}. The effective viscosity of the bacterial suspension decreases linearly when increasing the volume fraction $\phi_0$ of bacteria~\cite{saintillan_rheology_2018,lopez_turning_2015,Martinez2020,chui_rheology_2021}  
until a critical bacterial volume fraction $\phi_c$ is reached. Above this concentration, the activity of the bacteria is such that the viscosity becomes zero~\cite{lopez_turning_2015,Martinez2020,chui_rheology_2021}.
Our study demonstrates that the viscosity reduction observed when bacteria are added in a fluid can influence macroscopic flows and leads to flow instability such as Saffman--Taylor instability (also known as viscous fingering)  observed when a less viscous fluid displaces a more viscous fluid~\cite{Hill_1952,saffman_1958,Saffman1986ViscousFI,Homsy_1992, Chen1989,petitjean_maxworthy1_1996,Tabeling_Libchaber_1986,Tabeling_Zocchi_Libchaber_1987,Bensimon_Kadanoff_Liang_Shraiman_Tang_1986,Thome_Rabaud_Hakim_Couder_1989,lajeunesse_3d_1997, petitjeans_miscible_1999,Videbak_2020,keable_effect_2022}.
To demonstrate how the flow of bacterial suspensions can cause instabilities, we conducted experiments using the Saffman--Taylor geometry, which involves a  Hele-Shaw (HS) cell with parallel plates separated by a small aperture $H$. We identified the bacterial volume fraction and flow rate at which these instabilities occur. Additionally, we performed experiments in the quarter five-spot (QFS) configuration~\cite{petitjeans_miscible_1999,keable_effect_2022} to confirm the occurrence of multiple unstable fingers. These experiments also confirmed the significant role of the critical volume fraction $\phi_c$ coupled with a critical shear rate $\dot{\gamma}_c$ in triggering the instability.\\

\begin{figure*}[ht]
\centering
  \includegraphics[width=\textwidth]{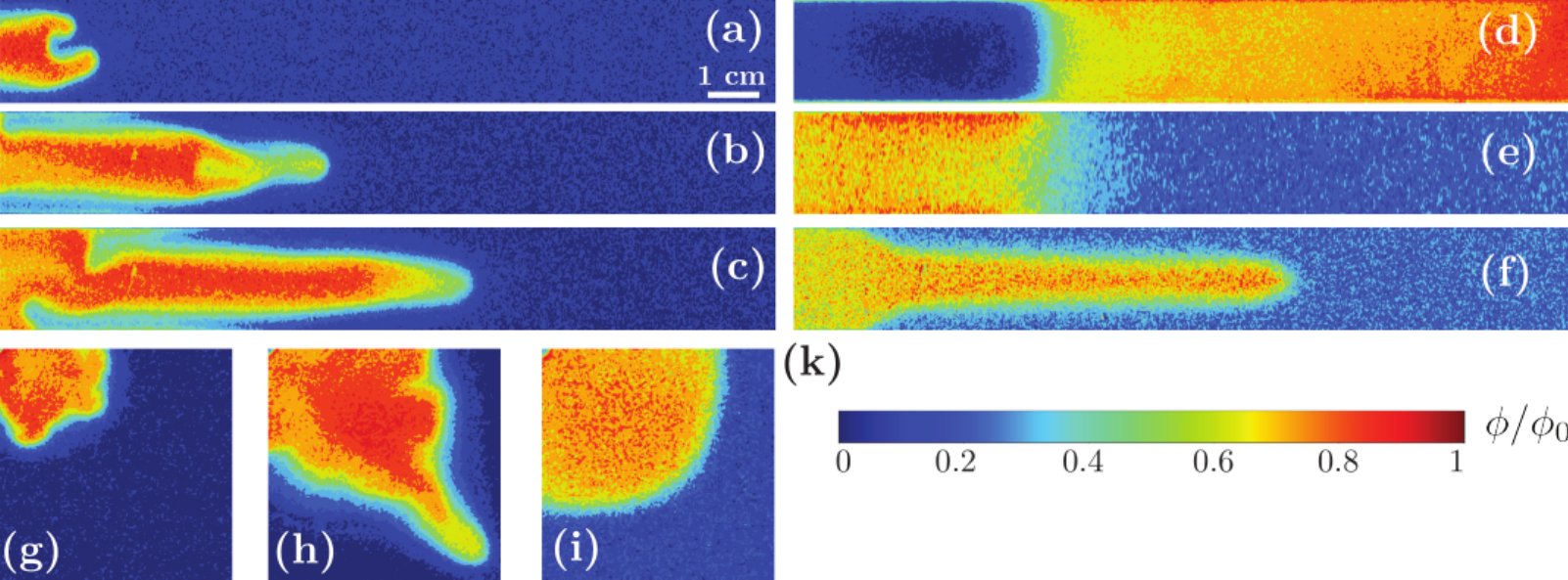}
  \caption{(a)-(c) Normalized bacterial volume fraction fields $\phi(x,y)/\phi_0$ for an experiment in HS cell of width $W = 2$~cm. The volume fraction of bacteria in the injected fluid is $\phi_0 \approx 0.825\% > \phi_c$. The images are taken 30 min apart. (d) Bacterial suspension of volume fraction $\phi_0 \approx 0.825\%$ displaced by the suspending fluid (reverse of (a)-(c)).(e) $\phi(x,y)/\phi_0$ when $\phi_0 \approx 0.11\% < \phi_c$. 
  (f) Experiment with Newtonian fluids with $M = 6.5$. For (a)-(f): $U \simeq 17~\mu$m.s$^{-1}$. (g)-(i) Normalized bacterial volume fraction fields $\phi(x,y)/\phi_0$ for experiments in the QFS geometry at $U \simeq 10~\mu$m.s$^{-1}$ with (g,h) $\phi_0 \approx 0.825\% > \phi_c$  and (i) $\phi_0 \approx 0.11\% < \phi_c$. (k) Color bar for all the fields. The scale bar for all figures is specified in bottom right of (a). 
  }
  \label{fig2}
\end{figure*}

\paragraph{\textbf{Methods.-} }
We used a wild-type strain of {\it Escherichia coli} (\textit{E. coli}) (RP437-YFP) that expresses yellow fluorescence protein. Bacteria were suspended in a motility buffer (MB) containing $1\%$ of PolyVinylPyrolydine (MB-1\%PVP) (SI~I). The concentration of bacteria is given in terms of volume fraction $\phi$, which was estimated from the OD (optical density) of the suspension with 1 OD corresponding to $\sim 8$.$10^5$ cells/$\mu$L. Assuming the volume of an individual bacterium to be $\sim 1.4~\mu$m$^{3}$ \cite{lopez_turning_2015}, this corresponds to a volume fraction $\phi \sim 0.11\%$. The average swimming velocity ($V_s$) and the rotational diffusivity ($D_R$) were obtained by tracking a dilute bacterial suspension under a microscope using a $10\times$ objective (SI~II). We observed $V_s = 15.0 \pm 1.2~\mu$m.s$^{-1}$ and $D_R= 0.09 \pm 0.02~$s$^{-1}$ when bacteria were suspended in MB-1\%PVP. 
The critical volume fraction was determined from rheological measurements conducted in a Couette rheometer~\cite{lopez_turning_2015,Martinez2020,chui_rheology_2021}. $\phi_c$ was taken between the last volume fraction with non-zero viscosity and the first volume fraction with zero viscosity. This gives $\phi_c = 0.66 \pm 0.05 \%$ for bacteria suspended in MB-1\%PVP. The rheological measurements (Fig.~\ref{fig1}(c)) also confirmed that at a shear rate lower than $0.1~$s$^{-1}$, the viscosity, $\eta_2(\phi_0)$, of the suspension is lower than the viscosity, $\eta_1$, of the same fluid without bacteria, confirming the behavior already reported in the literature~\cite{lopez_turning_2015,Martinez2020,chui_rheology_2021}.\\
The flow cells were created by pouring Polydimethylsiloxane (PDMS) onto master molds to produce replicas, which were then bonded to a thick glass plate through plasma activation (Fig.~\ref{fig1}(a)). PDMS is highly permeable to dioxygen, preventing the loss of bacterial motility.
The outlets were connected to a syringe pump that draws in the bacterial suspension from a reservoir connected to the channel's inlet at a fixed flow rate. 
The channel was first filled with the suspending fluid ($\eta_1=3.2\pm 0.1~$mPa.s) before the injection of the bacterial suspension. The channels were placed on top of a blue-light illumination table and a 16-bit CCD camera with an orange filter to visualize the progression of the bacterial concentration front. A calibration was performed to convert the fluorescence intensity integrated over the HS cell aperture into local bacterial volume fraction $\phi(x,y,t)$. The experiments were performed at different volume fractions $\phi_0$. The average flow velocity of the bacterial front $U$ was determined by performing flow displacement experiments at different flow rates using a dilute suspension  $\phi_0\simeq 0.066 \%<<\phi_c$~\cite{Ganesh_Douarche_Auradou_2023}. Fig.~\ref{fig1}(b) shows typical fields obtained with a spatial resolution of 500~$\mu$m for different bacterial volume fractions.\\

\begin{figure}[t]
\centering
\includegraphics[width=0.5\textwidth]{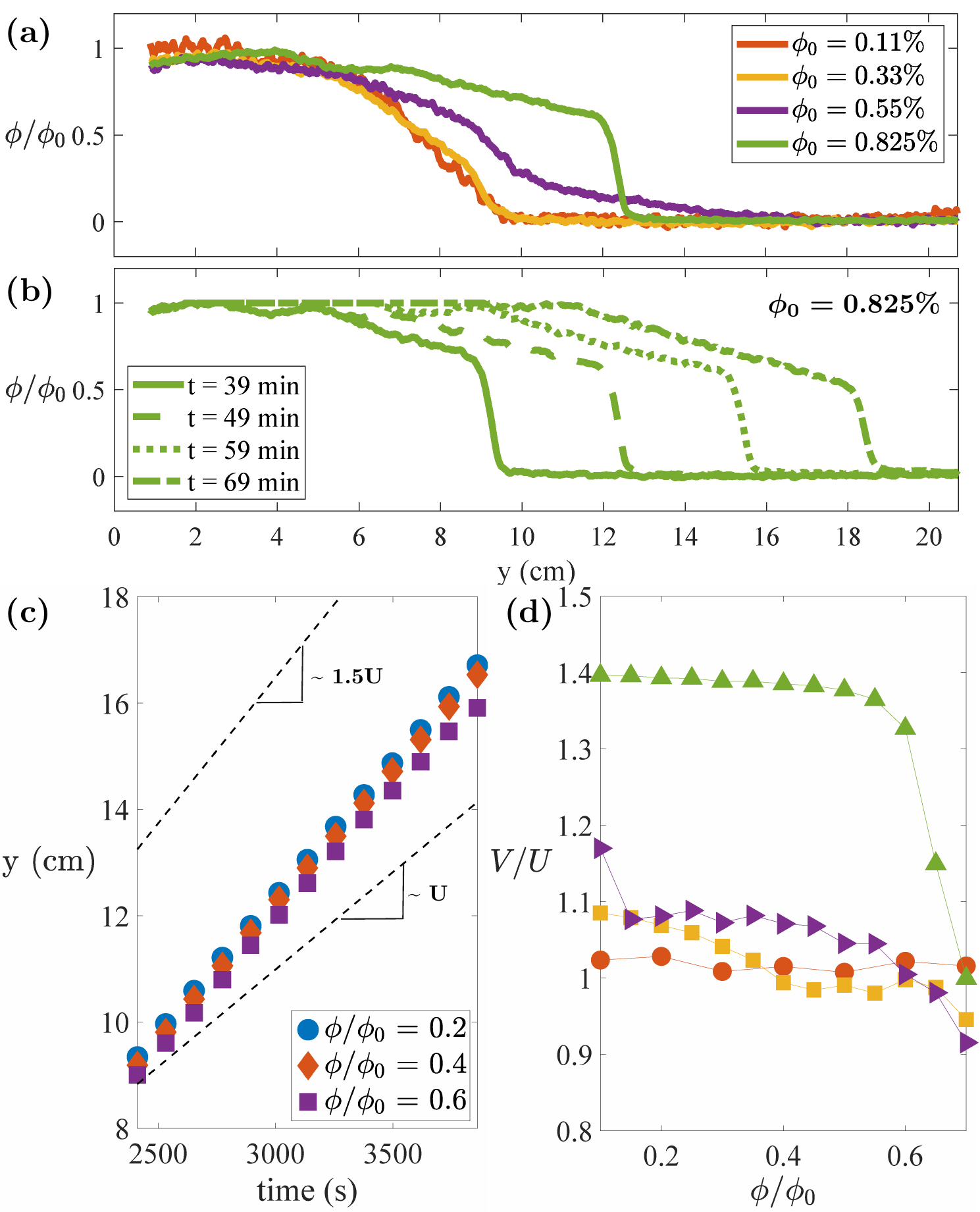} 
\caption{(a) Bacterial volume fraction profiles measured in an HS cell of width $W = 1$~cm plotted as a function of the distance, $y$, along the length of the cell. The profiles are obtained by averaging $\phi(x,y)$ in the direction of the cell width over a band of width $5~$mm passing through the center of the HS. The profiles are for a fixed time of $\sim 50$ min and for different volume fractions $\phi_0$.
(b) Profiles at different time measured for $\phi_0 \approx 0.825 \%~(>\phi_c)$.   
(c) $y$ position of the concentration profiles shown in (b) for three iso-volume fractions $\phi/\phi_0 = 0.2$~($\bigcirc$), 0.4 ($\lozenge$), and 0.6 ($\Box$) as a function of time. 
The slopes of these plots give $V(\phi/\phi_0)$. The dashed lines are a guide for the eyes. Here: $\phi_0\approx 0.825 \%$. (d) $V(\phi/\phi_0)/U$ as a function of $\phi/\phi_0$ for: $\phi_0 \approx 0.11\%(<\phi_c)$ ($\bigcirc$), $\phi_0 \approx 0.33\%(<\phi_c)$ ($\square$), $\phi_0 \approx 0.55\%(<\phi_c)$ ($\triangleright$) and  $\phi_0 \approx 0.825\%(>\phi_c)$ ($\triangle$). All experiments were done with $U \simeq 37~\mu$m.s$^{-1}$.}
  \label{fig3}
\end{figure}

\begin{figure}[t]
\centering
  \includegraphics[width=0.48\textwidth]{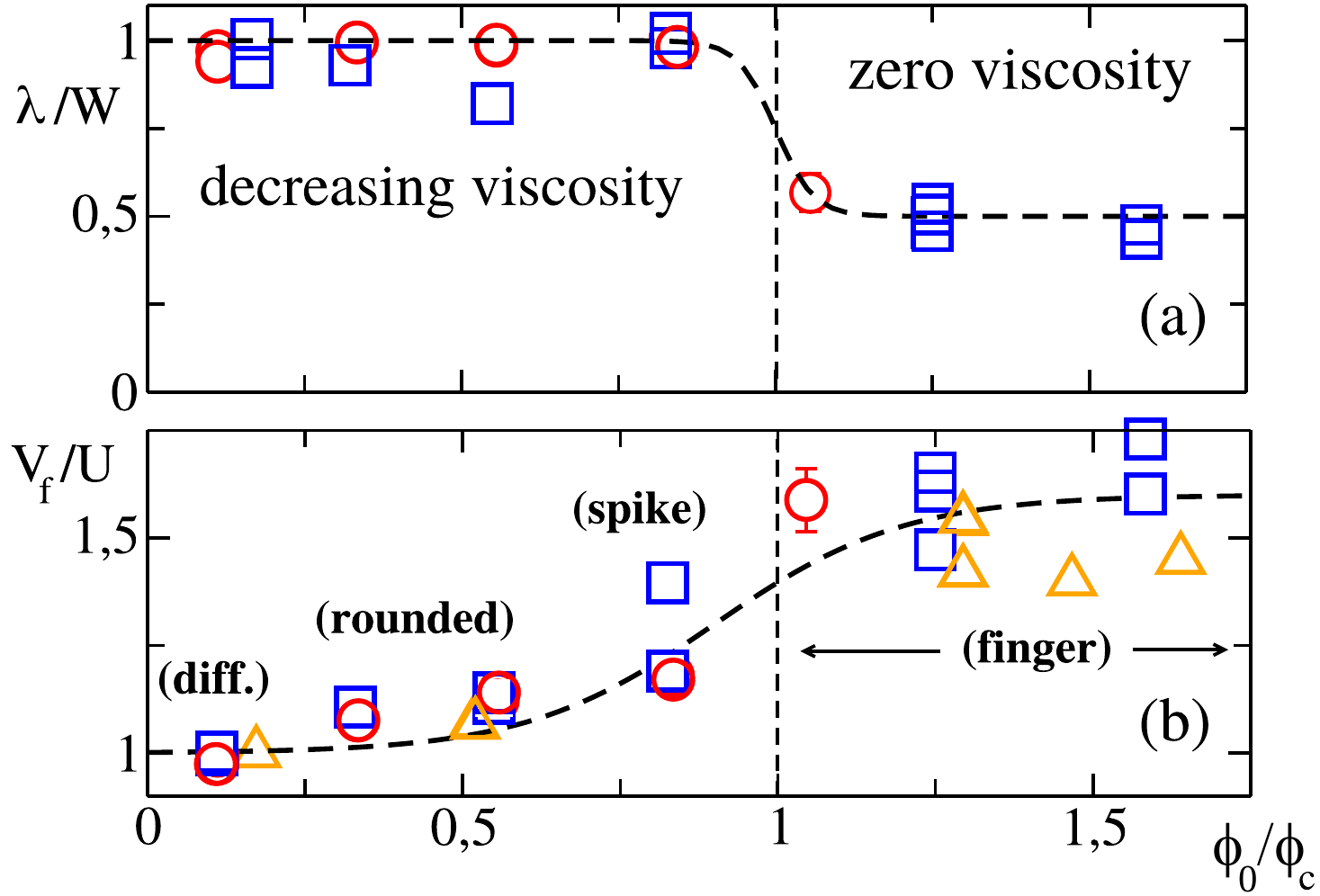}
  \caption{Normalized finger width $\lambda$ (a) and tip velocity $V_f$ (b) as a function of the bacterial volume fraction $\phi_0$ normalized with the critical volume fraction $\phi_c$. 
  $\square$ and $\bigtriangleup$: Bacteria in MB+$1\%$ PVP. $\bigcirc$: Bacteria in MB. $\bigtriangleup$: QFS experiments. $\square$ and $\bigcirc$: experiments in HS. 
  Dotted line, adjustment by: $\sim tanh(\frac{\phi_0}{\phi_c}-1)$.
  For the HS cell, $V_f=V(\phi/\phi_0=0.2)$ and $U$ is $\simeq 17$ or $37~\mu$m.s$^{-1}$. For the QFS experiments, $V_f$ is the distance between the input and the output, normalized by the time taken to reach the output. In this case, $U \simeq 10~\mu$m.s$^{-1}$.
  }
  \label{fig4}
\end{figure}

\paragraph{\textbf{Results.-} }
The images presented in Fig.~\ref{fig1}(b) provide clear evidence of the crucial role of the bacterial concentration in determining the shape of the displacing front. Based on the observations of the shapes, we identified four distinct fronts: diffusive, rounded, spike, and finger. 
The Fig.~\ref{fig2}(a-c) and \href{http://www.fast.u-psud.fr/~auradou/Active-Viscous-Fingering/}{Movie1} show the typical evolution of the unstable finger-like front. At the early stage of the instability (Fig.~\ref{fig2}(a)), the front showed two protrusions that were not present when the suspension contained a lower volume fraction of bacteria (Fig.~\ref{fig2}(e)). 
One of the protrusions expanded slightly ahead of its neighbor (Fig.~\ref{fig2}(b)) and led to the emergence of a single finger (Fig.~\ref{fig2}(c)). The back of the finger then expanded until it filled the entire width of the channel  (Fig.~\ref{fig2}(c)). The final finger shape observed in Fig.~\ref{fig2}(c), looked similar to the fronts obtained experimentally~\cite{saffman_1958,Tabeling_1986,Videbak_2020,baumgarten_viscous_2017,lajeunesse_3d_1997} and through simulations~\cite{rakotomalala_miscible_1997} with miscible or immiscible pair of Newtonian fluids. This was confirmed by conducting experiments with pairs of Newtonian fluids containing different quantities of PVP, the concentration of which was adjusted so that $M$, the viscosity ratio between the displaced and the displacing fluids, varied between 1 and 12. In these experiments, bacteria were replaced by fluorescein at a concentration of 0.02~g/L to visualize the progression of that fluid. Fig.~\ref{fig2}(f) shows one of the images obtained for $M \simeq 6.5$. The width of finger observed was similar to the one shown in Fig.~\ref{fig2}(c). 
Fig.~\ref{fig2}(d) shows an image obtained when the displaced and displacing fluids were swapped in the bacteria experiment ($\phi_0 > \phi_c$ here as well). Under this condition, the interface was stable, and no finger was observed (\href{http://www.fast.u-psud.fr/~auradou/Active-Viscous-Fingering/Movie2.avi}{Movie2}).
To confirm the role of the swimming activity of bacteria in the formation of the finger, and to eliminate the possibility of hydrodynamic instability caused by the alteration of fluid properties due to bacterial metabolic activity, we tested the effect of displacing the suspending fluid with the supernatant. We prepared a bacterial suspension with a volume fraction of $\phi_0~\approx~0.825\%$ and left the suspension undisturbed for three hours, after which we filtered the suspension to remove the bacteria. The fluid collected (supernatant) was then used to carry out the displacement experiment. We observed no emergence of instability in the front, indicating that any changes in the suspending fluid characteristics due to bacterial metabolic activity do not contribute to the instability observed.\\
 To determine whether the threshold $\phi_c$ for the onset of instability varied with geometry, we conducted experiments using the QFS geometry set-up. Again, experiments carried out at a very low bacterial volume fraction (Fig.~\ref{fig2}(i)) showed fields propagating similar to that of two Newtonian fluids with no viscosity contrast displacing each other(\href{http://www.fast.u-psud.fr/~auradou/Active-Viscous-Fingering/}{Movie3}). However, for $\phi_0~\approx~0.825\%$, the fields first showed multiple fingers, which appeared in the divergent part of the cell; these fingers then merged and flowed towards the outlet (Figs.~\ref{fig2}(g) and~\ref{fig2}(h)) \href{http://www.fast.u-psud.fr/~auradou/Active-Viscous-Fingering/}{Movie4}).\\
Fig.~\ref{fig3}(a) shows the volume fraction profiles of the four different regimes identified in Fig.~\ref{fig1}(b): 
at very low bacterial volume fraction, the profile exhibited a smooth and error-function-like pattern (red line: diffusive regime). At high bacterial volume fraction, a sharp shock front was observed (green line: finger regime) similar to that observed in viscous fingering experiments~\cite{Lajeunesse_Martin_Rakotomalala_Salin_Yortsos_1999,bischofberger_fingering_2014,Bischofberger2015}. In addition to these two regimes, we pointed out the rounded regime (yellow line) and the spike regime (purple line), where the latter is characterized by a spike of low bacterial volume fraction which traveled ahead of the front.\\
In the finger regime, a shock front is visible at all times (Fig.~\ref{fig3}(b)) and corresponds to a iso volume fraction range of $0 \lesssim \phi/\phi_0 \lesssim 0.5$. Fig.~\ref{fig3}(c) shows the temporal variation of three iso volume fraction positions. We observe that these positions $y(\frac{\phi}{\phi_0})$ vary linearly with time. 
The green symbols in Fig.~\ref{fig3}(d) show the normalized velocities $V(\frac{\phi}{\phi_0})/U$ obtained from the slopes of $y(\frac{\phi}{\phi_0})$ as a function of time. 
The region of volume fractions that corresponds to the sharp shock front ($0 \lesssim \frac{\phi}{\phi_0} \lesssim 0.5$) propagates at a normalized velocity $\sim 1.4$ faster than the average velocity of the fluid in a self-similar way.
When $\phi_0$ decreases to $\approx 0.55\%$ ($\triangleright$ in Fig.~\ref{fig3}(d)), the constant velocity plateau (the shock) disappeared. It is replaced by (i) a rapid drop of $V(\phi/\phi_0)/U$ for $\phi/\phi_0 \lesssim 0.15$ and (ii) a slower decrease for $\phi/\phi_0 \gtrsim 0.15$ towards $\sim 1$. The first sharp drop comes from a spike that contains a small number of bacteria that propagates ahead of the front. When $\phi_0$ decreases further to $\approx 0.33\%$ ($\square$ in Fig.~\ref{fig3}(d)), the spike disappears, and every iso-$\phi$'s of the profile spread in time with a $V(\phi/\phi_0)/U$ that decreases monotonically towards 1 when $\phi/\phi_0 \to 1$. This is the rounded regime. Finally, in the diffusive regime, for $\phi_0$ down to $\approx 0.11\%$ ($\bigcirc$ in Fig.~\ref{fig3}(d)), all the iso-$\phi$'s lines travel at a constant velocity close to $\sim 1$. 
These features are identical to those observed when a less viscous fluid displaces a more viscous fluid~\cite{petitjean_maxworthy1_1996,Lajeunesse_Martin_Rakotomalala_Salin_Yortsos_1999,bischofberger_fingering_2014,Bischofberger2015} and, the volume fraction of bacteria $\phi_0$ seems to play a similar role to that of the viscosity ratio $M$ for Newtonian fluids.\\
To test this idea, we characterized the shape of the front by its width $\lambda$ (SI~III) and its dynamics by the tip velocity $V_f$, as functions of $\phi_0/\phi_c$. The $\square$ in Fig.~\ref{fig4}(a) illustrate the variation of the finger width $\lambda$ normalized by the channel width $W$. For $\phi_0 \lesssim \phi_c$, the normalized finger width is close to 1, and the instability is absent. The normalized width decreases rapidly for $\phi \gtrsim \phi_c$ and tends towards half of the cell's width, emphasizing the emergence of fingers. 
To support this observation, experiments were conducted using bacterial suspensions of different critical volume fractions $\phi_c$. In MB, for instance, the bacteria swim at a slower speed ($V_s=10.6 \pm 1.2~\mu$m.s$^{-1}$), and the rotational diffusion coefficient is higher ($D_R = 0.13 \pm 0.02$~s$^{-1}$) compared to MB+$1\%$ PVP. As a result, a larger quantity of bacteria $\phi_c \approx 0.99\%$ is needed to reach the zero viscosity regime. $\bigcirc$ in Fig.~\ref{fig4}(a) shows the value of $\lambda/W$ as a function of $\phi_0/\phi_c$ for a suspension of bacteria in MB displacing MB. We observed again that $\lambda/W$ decreases to $\sim 0.5$ for $\phi \gtrsim \phi_c$, which corresponds to the fingering instability emergence. 
In the same way, we observe that at low $\phi_0$, the normalized tip velocities are close to 1 (Fig.~\ref{fig4}(b)), and fingers are absent. 
The same analyses were conducted on experiments with Newtonian fluid pairs. The results shown in SI~IV indicate that when the viscosity contrast is approximately 3, the displacement front becomes unstable \cite{lajeunesse_3d_1997,bischofberger_fingering_2014}. Referring to Fig.~\ref{fig1}(c), we observe that the normalized volume fraction $\phi_0/\phi_c$ needed to achieve this viscosity ratio is approximately 0.5. Experiments conducted with this volume fraction demonstrate a stable front, which is equal in width to that of the cell (Fig.~\ref{fig4}(a)), moving at the average velocity of the fluid (Fig.~\ref{fig4}(b)). This indicated that the instability observed with bacterial suspensions is specific to a threshold linked to the zero-viscosity regime. It emphasizes the importance of using a suspension with a volume fraction greater than $\phi_c$ to observe the emergence of an instability.\\
Another difference with Newtonian fluids is that in the unstable configuration, increasing the flow velocity leads to the formation of multiple fingers ~\cite{Taylor1961,Tabeling_Libchaber_1986,malhotra_experimental_2015,Bischofberger2015}. Surprisingly, with bacteria, we observed the finger regime to disappear when the velocity was increased to 100~$\mu$m.s$^{-1}$.
One way to explain this is through the unique rheological behavior of the bacterial suspensions~\cite{lopez_turning_2015, Martinez2020, chui_rheology_2021}.  
At low shear rates, the suspensions show a low-viscosity Newtonian plateau, whose value (represented by $\Diamond$ in Fig.~\ref{fig1}(c)) changes with the bacterial volume fraction. When the shear rates are higher than $\dot{\gamma}_c$, the suspensions exhibit a second higher Newtonian plateau with a viscosity close to that of the suspending fluid  (represented by $\bigcirc$ in Fig.~\ref{fig1}(c)).
The crossover between the two plateaus occurs for  $\dot{\gamma_c} \sim 4 D_R \sim 0.4$~s$^{-1}$~\cite{chui_rheology_2021}. If we assume a parabolic velocity profile in the aperture of the HS cell, the flow can be characterized by an average shear rate $\langle \dot{\gamma}\rangle=\frac{4U}{H}$. For velocities below $37~\mu$m.s$^{-1}$, $\langle \dot{\gamma}\rangle$ is less than $0.4$~s$^{-1}$ \textit{i.e.} below $\dot{\gamma}_c$. The fluid dynamics within the HS cell are affected by viscosity reduction, resulting in an effective low-viscosity fluid and the formation of a finger. For $U \sim 100~\mu$m.s$^{-1}$, $\langle \dot{\gamma}\rangle$ is $\sim 1~$s$^{-1}$ higher than $\dot{\gamma}_c$ which leads to the second Newtonian plateau. The difference in viscosity between the displaced and injected fluids is then not enough to cause instability.\\
\paragraph{\textbf{Discussion and conclusions.-}}
We observed that introducing swimming bacteria into a flowing fluid can destabilize the interface between the bacterial front and the displaced fluid, resulting in a new interfacial dynamics. This instability was observed when both $\phi_0 \gtrsim \phi_c$ and $\langle \dot{\gamma}\rangle \lesssim \dot{\gamma}_c$ for which the bacterial suspension effective viscosity is zero. For lower $\phi_0$, the swimming activity of the bacteria seems too weak to cause any such instability but leads to some alterations of the density profile. 
Another interesting observation is the emergence of vorticity in our experiments in a QFS geometry, which may be related to collective motion in the suspension. This observation supports the finding of a sudden increase in the length of velocity correlations, which reaches dimensions comparable to the cell opening when $\phi_0 > \phi_c$~\cite{Martinez2020}. Future research may go beyond correlation to show their connection.
We think our findings are crucial in situations where bacteria travel through tight spaces such as pores, fractures, or slits. They can

be useful for developing technology based on biology where reducing fluid viscosity is important, for example, in microbial-enhanced oil recovery (MEOR).\\
By showing that activity at the individual particle's scale can control the fluid's effective property and influence fluid flow, our results also significantly contribute to the already rich field of active matter and open new paths for further research.\\

\paragraph{\textbf{Acknowledgements}}
We thank A. Gargasson, A. Aubertin, C. Manquest,  L. Auffray, and R. Pidoux for experimental help, and H. Stone, F. Moisy, M. Jarrahi, G. Dietze and F. Doumenc for fruitful discussions.
This work is supported by the French National Research Agency (ANR) through the "Laboratoire d'Excellence Physics Atom Light Mater" (LabEx PALM) as part of the "Investissements d'Avenir" program (ANR-10-LABX-0039). H.A. acknowledged the support of CNRS 80~$|$~PRIME through the project RootBac.

\bibliographystyle{apsrev4-1}

\end{document}